\documentclass[12pt,a4paper]{JHEP3}

\usepackage[cmtip,arrow]{xy}
\usepackage{pb-diagram,pb-xy}

\newlength{\vshift}
\newlength{\hshift}
\usepackage{amssymb,amsopn}

\def\nn{\nonumber }
\def\la{\lambda}

\def\be{\beta}
\def\al{\alpha}
\def\si{\sigma}

\def\ds{\stackrel{\star}{,}}

\def\p{\partial}

\def\lb{\lbrack}
\def\rb{\rbrack}
\def\p{\partial}

\def\KK{K^{m}_{\ \ l}K^{*nl}}
\def\KKt{K^{ab}K_{ab}}
\def\KKtc{K^{*cd}K_{cd}^*}
\def\KKi{K^m_{\ \ a}K^{*n}_{\ \ \ b}}

\title{Field Theory on Nonanticommutative Superspace}

\author{
Marija Dimitrijevi\' c,\\
INFN Gruppo Collegato di Alessandria,\\
Via Bellini 25/G 15100 Alessandria, Italy and\\

Faculty of Physics, University of Belgrade, \\
P. O. Box 368, 11001 Belgrade, Serbia
\\E-mail:
\email{dimitrij@to.infn.it} and \email{dmarija@phy.bg.ac.yu}}

\author{
Voja Radovanovi\' c
\\Faculty of Physics, University of Belgrade, \\
P. O. Box 368, 11001 Belgrade, Serbia
\\E-mail:
\email{rvoja@phy.bg.ac.yu}}

\author{ \begin{tabular}{|c|}\hline
Julius Wess\\ \hline
\end{tabular}
   \\    Arnold Sommerfeld Centre for Theoretical Physics\\
Universit\"at M\"unchen, Fakult\"at f\"ur Physik\\
Theresienstr.\ 37, 80333 M\"unchen, Germany,\\

Max-Planck-Institut f\"ur Physik\\
F\"ohringer Ring 6, 80805 M\"unchen, Germany and\\

Universit\"at Hamburg, II Institut f\"ur Theoretische Physik\\
and DESY, Hamburg\\
        Luruper Chaussee 149, 22761 Hamburg, Germany\\  }

\abstract{ We discuss a deformation of the Hopf algebra of
supersymmetry (SUSY) transformations based on the special choice of
twist. As usual, algebra itself remains unchanged, but the
comultiplication changes. This leads to the deformed Leibniz rule
for SUSY transformations. Superfields are elements of the algebra
of functions of the usual supercoordinates. Elements of this
algebra are multiplied by using the $\star$-product which is
noncommutative, hermitian and finite when expanded in power series
of the deformation parameter. Chiral fields are no longer a
subalgebra of the algebra of superfields. One possible deformation
of the Wess-Zumino action is proposed and analyzed in detail.
Differently from most of the literature concerning this
subject, we work in Minkowski space-time.}

\keywords{ supersymmetry, twist, non(anti)commutative space,
deformed Wess-Zumino model}

\begin{document}

\section{Introduction}

It is well known that Quantum Field Theory (QFT) encounters
problems at very high energies and very short distances. This
suggests that the structure of space-time has to be modified
at these scales. One possibility to modify the structure of space-time is to
deform the usual commutation relations between coordinates; this gives a noncommutative (NC) space \cite{noncommspace}. Different models of noncommutativity were
discussed in the literature. One of the simplest
examples is the $\theta$-deformed or canonically deformed
space-time \cite{theta} with
\begin{equation}
\lb x^m, x^n \rb = i\theta^{mn} .\label{thetacom}
\end{equation}
Here $\theta^{mn}$ is a constant antisymmetric matrix. Gauge
theories were defined and analyzed in details in this framework \cite{jssw}. Also,
a deformed Standard Model was formulated \cite{smtheta} and
renormalizability properties of field theories on this space are
subject of many papers \cite{ren}.

More complicated deformations of space-time, such as
$\kappa$-deformation \cite{luk} and $q$-deformation \cite{abe}  were
also discussed in the literature.

In order to understand the physics at very small scales better,
in recent years attempts were made to combine
supersymmetry with noncommutativity. In \cite{luksusy} the authors
combine SUSY with the $\kappa$-deformation
of space-time, while in \cite{MWsusy} SUSY is combined with the
canonical deformation of space-time. In series of papers
\cite{nonantisusy}, \cite{Seiberg}, \cite{Ferrara}
a version of non(anti)commutative superspace is defined and analyzed.  The
anticommutation relations between the fermionic coordinates are
modified in the following way
\begin{equation}
\{ \theta^\alpha \ds \theta^\beta\} =C^{\alpha\beta} , \quad \{
\bar{\theta}_{\dot\alpha} \ds \bar{\theta}_{\dot\beta}\} = \{
\theta^\alpha \ds \bar{\theta}_{\dot\alpha}\} = 0\ ,
\label{eucliddef}
\end{equation}
where $C^{\alpha\beta} = C^{\beta\alpha}$ is a complex, constant
symmetric matrix. Such deformation is well defined only in Euclidean
space where undotted and dotted spinors are not related by the
usual complex conjugation. Note that the chiral coordinates
$y^m=x^m+i\theta\sigma^m\bar\theta$ commute in this setting.

In \cite{Seiberg} the notion of chirality is preserved, i.e.
the deformed product of two chiral superfields is again a chiral
superfield. On the other hand, one half of $N=1$ supersymmetry is broken and this is
the so-called $N=1/2$ supersymmetry. Another type of deformation is
introduced in \cite{Ferrara}. There the product of two chiral
superfields is not a chiral superfield but the model is invariant
under the full supersymmetry. The Hopf algebra of SUSY
transformations is deformed by using the twist approach in
\cite{chiralstar}. Examples of
deformation that introduce nontrivial commutation relations between
chiral and fermionic coordinates are discussed in \cite{14}. Some
consequences of nontrivial (anti)commutation relations on statistics and S-matrix are analyzed in \cite{15}.

In this paper we apply a twist to deform the Hopf algebra of
SUSY transformations. However, our choice of the twist is
different from that in \cite{chiralstar} since we want to work in
Minkowski space-time. As undotted and dotted spinors are related
by the usual complex conjugation, we obtain
\begin{equation}
\{ \theta^\alpha \ds \theta^\beta\} = C^{\alpha\beta} , \quad \{
\bar{\theta}_{\dot\alpha} \ds \bar{\theta}_{\dot\beta}\} =
\bar{C}_{\dot{\alpha}\dot{\beta}}, \quad \{ \theta^\alpha \ds
\bar{\theta}_{\dot\alpha}\} = 0\ , \label{minkdef}
\end{equation}
with $\bar{C}_{\dot{\alpha}\dot{\beta}} = (C_{\alpha\beta})^*$.
Our main goal is the formulation and analysis of the deformed
Wess-Zumino Lagrangian.

The paper is organized as follows: In section 2 we review the
undeformed supersymmetric theory to establish the notation and
then rewrite it by using the language of Hopf algebras. We follow the notation of
\cite{wessbook}. By twisting the Hopf algebra of SUSY
transformations, a Hopf algebra of deformed SUSY transformations
is obtained in section 3. As the algebra itself remains undeformed,
the full $N=1$ SUSY is preserved. On the other hand, the comultiplication
changes and that leads to a deformed Leibniz rule. As a
consequence of the twist, a $\star$-product is introduced on the
algebra of functions of supercoordinates. Sections 4 and 5 are devoted to the
construction of a deformed Wess-Zumino Lagrangian. Since our
choice of the twist implies that the $\star$-product of chiral superfields is
not a chiral superfield we have to use (anti)chiral projectors to
project irreducible components of such $\star$-products. In the section
6 the auxiliary fields are integrated out and the expansion in the
deformation parameter of the "on-shell" action is given. Some consequences of applying the twist on the Poincar\' e invariance are discussed in the section 7. Two examples of how to apply the deformed Leibniz rule when transforming $\star$-products of fields are given. Finally, we end the paper with some short comments and conclusions.

\section{Undeformed SUSY transformations}

The undeformed superspace is generated by $x$, $\theta$ and
$\bar{\theta}$ coordinates which fulfill
\begin{eqnarray}
\lb x^m, x^n \rb &=& \lb x^m,
\theta^\alpha \rb = \lb x^m, \bar{\theta}_{\dot\alpha} \rb = 0 ,\nonumber\\
\{ \theta^\alpha , \theta^\beta\} &=& \{ \bar{\theta}_{\dot\alpha}
, \bar{\theta}_{\dot\beta}\} = \{ \theta^\alpha ,
\bar{\theta}_{\dot\alpha}\} = 0 , \label{undefsupsp}
\end{eqnarray}
with $m=0,\dots 3$ and $\alpha, \beta =1,2$. These coordinates we
call the supercoordinates, to $x^m$ we refer as to bosonic and to
$\theta^\alpha$ and $\bar{\theta}_{\dot\alpha}$ we refer as
to fermionic coordinates. Also, $x^2= x^m x_m =
-(x^0)^2 + (x^1)^2 + (x^2)^2 + (x^3)^2$, that is we work in
Minkowski space-time with the metric $(-,+,+,+)$.

Every function of the supercoordinates can be expanded in power series
in $\theta$ and $\bar{\theta}$. Superfields form a subalgebra of
the algebra of functions on the superspace. For a general superfield
$F(x,\theta, \bar{\theta})$ the expansion in $\theta$ and
$\bar{\theta}$ reads
\begin{eqnarray}
F(x, \theta, \bar{\theta}) &=&\hspace*{-2mm} f(x) + \theta\phi(x)
+ \bar{\theta}\bar{\chi}(x)
+ \theta\theta m(x) + \bar{\theta}\bar{\theta} n(x) + \theta\sigma^m\bar{\theta}v_m\nonumber\\
&& + \theta\theta\bar{\theta}\bar{\lambda}(x) + \bar{\theta}\bar{\theta}\theta\varphi(x)
+ \theta\theta\bar{\theta}\bar{\theta} d(x) .\label{F}
\end{eqnarray}
All higher powers of $\theta$ and $\bar{\theta}$ vanish since
these coordinates are Grassmanian.

Under the infinitesimal SUSY transformations a general superfield transforms as
\begin{equation}
\delta_\xi F = \big(\xi Q + \bar{\xi}\bar{Q} \big) F, \label{susytr}
\end{equation}
where $\xi$ and $\bar{\xi}$ are constant anticommuting parameters
and $Q$ and $\bar{Q}$ are SUSY generators
\begin{eqnarray}
Q_\alpha &=& \p_\alpha
- i\sigma^m_{\ \alpha\dot{\alpha}}\bar{\theta}^{\dot{\alpha}}\p_m, \label{q}\\
\bar{Q}^{\dot{\alpha}} &=& \bar{\p}^{\dot{\alpha}} -
i\theta^\alpha \sigma^m_{\
\alpha\dot{\beta}}\varepsilon^{\dot{\beta}\dot{\alpha}}\p_m
.\label{barq}
\end{eqnarray}
Using the expansion (\ref{F}) one can calculate the transformation
law of the component fields
\begin{eqnarray}
\delta_\xi f &=& \xi^\alpha \phi_\alpha
+ \bar{\xi}_{\dot\alpha}\bar{\chi}^{\dot\alpha}, \label{susytrf} \\
\delta_\xi \phi_\alpha &=& 2\xi_\alpha m
+ \sigma^m_{\ \alpha\dot{\alpha}}\bar{\xi}^{\dot\alpha}\big( v_m + i (\p_m f) \big), \label{susytrphi} \\
\delta_\xi \bar{\chi}^{\dot\alpha} &=& 2\bar{\xi}^{\dot\alpha}n
+ \bar{\sigma}^{m\dot{\alpha}\alpha}\xi_\alpha\big( -v_m + i (\p_m f) \big), \label{susytrchi} \\
\delta_\xi m &=&\bar{\xi}_{\dot\alpha}\bar{\lambda}^{\dot\alpha}
+ \frac{i}{2}\bar{\xi}_{\dot\alpha}\bar{\sigma}^{m\dot{\alpha}\alpha}(\p_m \phi_\alpha) , \label{susytrm} \\
\delta_\xi n &=& \xi^\alpha \varphi_\alpha
+ \frac{i}{2}\xi^\alpha\sigma^m_{\ \alpha\dot{\alpha}}(\p_m \bar{\chi}^{\dot\alpha}), \label{susytrn} \\
\sigma^m_{\ \alpha\dot{\alpha}}\delta_\xi v_m &=&
-i(\p_m\phi_\alpha)\xi^\beta\sigma^m_{\ \beta\dot{\alpha}} +
2\xi_\alpha\bar{\lambda}_{\dot\alpha} + i\sigma^m_{\
\alpha\dot{\beta}}\bar{\xi}^{\dot\beta}(\p_m
\bar{\chi}_{\dot\alpha})
+ 2\varphi_\alpha\bar{\xi}_{\dot\alpha} , \label{susytrvm} \\
\delta_\xi \bar{\lambda}^{\dot\alpha} &=& 2\bar{\xi}^{\dot\alpha}d
+ i\bar{\sigma}^{l\dot{\alpha}\alpha}\xi_\alpha(\p_l m)
+ \frac{i}{2}\bar{\sigma}^{l\dot{\alpha}\alpha}\sigma^m_{\ \alpha\dot{\beta}}\bar{\xi}^{\dot\beta}(\p_m v_l) , \label{susytrlambda} \\
\delta_\xi \varphi_\alpha &=& 2\xi_\alpha d + i\sigma^l_{\
\alpha\dot{\alpha}}\bar{\xi}^{\dot\alpha}(\p_l n)
-\frac{i}{2}\sigma^l_{\ \alpha\dot{\alpha}}\bar{\sigma}^{m\dot{\alpha}\beta}\xi_\beta(\p_m v_l) , \label{susytrvarphi} \\
\delta_\xi d &=& \frac{i}{2}\xi^\alpha \sigma^m_{\
\alpha\dot{\alpha}}(\p_m \bar{\lambda}^{\dot\alpha})
-\frac{i}{2}(\p_m \varphi^\alpha)\sigma^m_{\
\alpha\dot{\alpha}}\bar{\xi}^{\dot\alpha} . \label{susytrd}
\end{eqnarray}
Transformations (\ref{susytr}) close in the algebra
\begin{equation}
[\delta_\xi, \delta_\eta] = -2i(\eta\sigma^m \bar{\xi} - \xi\sigma^m \bar{\eta})\p_m. \label{xietaalg}
\end{equation}

We next consider the product of two superfields defined as
\begin{equation}
F\cdot G = \mu \{ F\otimes G\}, \label{defpr}
\end{equation}
where the bilinear map $\mu$ maps the tensor product to the space of functions. The
transformation law of the product (\ref{defpr}) is given by
\begin{eqnarray}
\delta_\xi (F\cdot G) &=& \big(\xi Q + \bar{\xi}\bar{Q} \big) (F\cdot G), \nonumber\\
&=& (\delta_\xi F)\cdot G + F\cdot(\delta_\xi G).
\label{undefLrule}
\end{eqnarray}
The first line tells us that the product of two superfields is a
superfield again. The second line is the usual Leibniz rule.

All these properties we sumarise in the language of Hopf algebras
\cite{abe}, which will be useful when we introduce a deformation
of the superspace. The Hopf algebra of undeformed SUSY
transformations is given by
\begin{itemize}
\item
algebra
\begin{eqnarray}
[\delta_\xi, \delta_\eta] = -2i(\eta\sigma^m \bar{\xi} - \xi\sigma^m
\bar{\eta})\p_m, \quad [\p_m, \p_n] = [\p_m, \delta_\xi] =0. \nonumber
\end{eqnarray}

\item
coproduct
\begin{eqnarray}
&&\Delta (\delta_\xi) = \delta_\xi \otimes 1 + 1\otimes
\delta_\xi, \quad \Delta \p_m = \p_m \otimes 1 + 1\otimes \p_m.
\label{Hopfdeltaxi}
\end{eqnarray}

\item
counit and antipode
\begin{equation}
\varepsilon(\delta_\xi)= \varepsilon(\p_m) =0, \quad S(\delta_\xi)
= -\delta_\xi, \quad S(\p_m) = -\p_m. \nonumber
\end{equation}

\end{itemize}
In the language of generators $Q_\alpha$ and
$\bar{Q}_{\dot{\alpha}}$ this Hopf algebra reads
\begin{itemize}
\item
algebra
\begin{eqnarray}
&& \{ Q_\alpha , Q_\beta\} = \{ \bar{Q}_{\dot{\alpha}} , \bar{Q}_{\dot{\beta}}\} = 0, \quad
\{ Q_\alpha , \bar{Q}_{\dot{\beta}} \} = 2i\sigma^m_{\ \alpha\dot{\beta}}\p_m ,\nonumber\\
&& \hspace*{1cm}[\p_m, \p_n] = [\p_m, Q_\alpha] = [\p_m, \bar{Q}_{\dot{\alpha}}] =0 .\label{Qcom}
\end{eqnarray}

\item
coproduct
\begin{eqnarray}
&&\Delta Q_\alpha = Q_\alpha \otimes 1 + 1\otimes Q_\alpha, \quad
\Delta\bar{Q}_{\dot{\alpha}} = \bar{Q}_{\dot{\alpha}} \otimes 1 + 1\otimes \bar{Q}_{\dot{\alpha}}, \nonumber \\
&&\hspace*{2cm} \Delta \p_m = \p_m \otimes 1 + 1\otimes \p_m. \label{coprQ}
\end{eqnarray}

\item
counit and antipode
\begin{eqnarray}
&& \varepsilon(Q_\alpha)=\varepsilon(\bar{Q}_{\dot{\alpha}})=
\varepsilon(\p_m) =0, \nonumber \\
&& S(Q_\alpha) = -Q_\alpha, \quad S(\bar{Q}_{\dot{\alpha}}) =
-\bar{Q}_{\dot{\alpha}}, \quad S(\p_m) = -\p_m
.\label{undefcounitantipod}
\end{eqnarray}

\end{itemize}

\section{Twisted SUSY transformations}

As in \cite{defgt} we introduce the deformed SUSY transformations by
twisting the usual Hopf algebra (\ref{Hopfdeltaxi}). For the twist
$\cal{F}$ we choose
\begin{equation}
{\cal F} = e^{\frac{1}{2}C^{\alpha\beta}\p_\alpha \otimes\p_\beta
+ \frac{1}{2}\bar{C}_{\dot{\alpha}\dot{\beta}}\bar{\p}^{\dot{\alpha}}\otimes\bar{\p}^{\dot{\beta}} } ,\label{twist}
\end{equation}
with $C^{\alpha\beta} = C^{\beta\alpha}$ a complex constant matrix. Note that
$C^{\alpha\beta}$ and $\bar{C}^{\dot{\alpha}\dot{\beta}}$ are
related by the usual complex conjugation. It was shown in \cite{twist2} that
(\ref{twist}) satisfies all the requirements for a twist \cite{chpr}. The twisted Hopf algebra of SUSY transformation now reads
\begin{itemize}

\item
algebra
\begin{eqnarray}
\{ Q_\alpha , Q_\beta\} &=& \{ \bar{Q}_{\dot{\alpha}} ,
\bar{Q}_{\dot{\beta}}\} = 0, \quad
\{ Q_\alpha , \bar{Q}_{\dot{\beta}} \} = 2i\sigma^m_{\ \alpha\dot{\beta}}\p_m ,\nonumber\\
\lbrack \p_m, \p_n \rbrack &=& [\p_m, \p_\alpha] = [\p_m, \bar{\p}_{\dot{\beta}}] = [\p_m, Q_\alpha] = [\p_m, \bar{Q}_{\dot{\alpha}}] =0 , \nonumber\\
\{ \p_\alpha , \p_\beta\} &=& \{ \p_\alpha , \bar{\p}_{\dot{\beta}} \}
= \{ \bar{\p}_{\dot{\alpha}} , \bar{\p}_{\dot{\beta}} \} = \{ \p_\alpha , Q_\beta\}
= \{ \bar{\p}_{\dot{\alpha}} , \bar{Q}^{\dot{\beta}} \} = 0, \label{defQcom}\\
\{ \p_\alpha , \bar{Q}^{\dot{\alpha}} \} &=& -i\sigma^m_{\ \alpha\dot{\beta}}\varepsilon^{\dot{\beta}\dot{\alpha}}\p_m, \quad
\{ \bar{\p}_{\dot{\alpha}} , Q_\alpha \} = -i\sigma^m_{\ \alpha\dot{\alpha}}\p_m.
\nonumber
\end{eqnarray}

\item
coproduct
\begin{eqnarray}
\Delta_{\cal F}(Q_\alpha) &=& {\cal F}
\Big( Q_\alpha \otimes 1 + 1\otimes Q_\alpha \Big){\cal F}^{-1} \nonumber\\
&=&Q_\alpha \otimes 1 + 1\otimes Q_\alpha\nonumber\\
&& -\frac{i}{2}\bar{C}_{\dot{\alpha}\dot{\beta}} \Big( \sigma^m_{\
\alpha\dot{\gamma}}\varepsilon^{\dot{\gamma}\dot{\alpha}}\p_m
\otimes
\bar{\p}^{\dot{\beta}} + \bar{\p}^{\dot{\alpha}} \otimes
\sigma^m_{\ \alpha\dot{\gamma}}\varepsilon^{\dot{\gamma}\dot{\beta}}\p_m \Big) , \nonumber\\
\Delta_{\cal F}(\bar{Q}_{\dot{\alpha}}) &=&
\bar{Q}_{\dot{\alpha}} \otimes 1 + 1\otimes \bar{Q}_{\dot{\alpha}} \label{defcopr}\\
&& + \frac{i}{2}C^{\alpha\beta}\Big(
\sigma^m_{\ \alpha\dot{\alpha}}\p_m\otimes\p_\beta + \p_\alpha\otimes \sigma^m_{\ \beta\dot{\alpha}}\p_m \Big) ,\nonumber\\
\Delta \p_m &=& \p_m \otimes 1 + 1\otimes \p_m, \nonumber\\
\Delta \p_\alpha &=& \p_\alpha \otimes 1 + 1\otimes \p_\alpha, \quad
\Delta \bar{\p}^{\dot{\alpha}} = \bar{\p}^{\dot{\alpha}} \otimes 1 + 1\otimes \bar{\p}^{\dot{\alpha}}. \nonumber
\end{eqnarray}

\item
counit and antipode
\begin{eqnarray}
\varepsilon(Q_\alpha) &=& \varepsilon(\bar{Q}_{\dot{\alpha}})=
\varepsilon(\p_m) = \varepsilon(\p_\alpha) = \varepsilon( \bar{\p}^{\dot{\alpha}} ) = 0, \nonumber \\
S(Q_\alpha) &=& -Q_\alpha, \quad S(\bar{Q}_{\dot{\alpha}}) =
-\bar{Q}_{\dot{\alpha}}, \nonumber\\
S(\p_m) &=& -\p_m, \quad S(\p_\alpha) = -\p_\alpha, \quad
S(\bar{\p}^{\dot{\alpha}} )= - \bar{\p}^{\dot{\alpha}}. \label{defcounitantipod}
\end{eqnarray}

\end{itemize}
Note that only the coproduct is changed, while the algebra stays the
same as in the undeformed case. This means that the full
supersymmetry is preserved. Also note that in order for the comultiplication for $Q_\alpha$ and $\bar{Q}_{\dot{\alpha}}$ to close in the algebra, we had to enlarge the algebra by introducing the fermionic derivatives $\p_\alpha$ and $\bar{\p}_{\dot{\alpha}}$.

The inverse of the twist (\ref{twist})
\begin{equation}
{\cal F}^{-1} = e^{-\frac{1}{2}C^{\alpha\beta}\p_\alpha
\otimes\p_\beta -
\frac{1}{2}\bar{C}_{\dot{\alpha}\dot{\beta}}\bar{\p}^{\dot{\alpha}}\otimes\bar{\p}^{\dot{\beta}}
} ,\label{invtwist}
\end{equation}
defines a new product on the algebra of functions of
supercoordinates called the $\star$-product. For two arbitrary
superfields $F$ and $G$ the $\star$-product is defined as follows
\begin{eqnarray}
F\star G &=& \mu_\star \{ F\otimes G \} \nonumber\\
&=& \mu \{ {\cal F}^{-1}\, F\otimes G\} \nonumber\\
&=& \mu \{ e^{-\frac{1}{2}C^{\alpha\beta}\p_\alpha \otimes\p_\beta
-
\frac{1}{2}\bar{C}_{\dot{\alpha}\dot{\beta}}\bar{\p}^{\dot{\alpha}}
\otimes\bar{\p}^{\dot{\beta}}} F\otimes G \} \label{starpr}\\
&=& F\cdot G - \frac{1}{2}(-1)^{|F|}C^{\alpha\beta}(\p_\alpha
F)\cdot(\p_\beta G) -
\frac{1}{2}(-1)^{|F|}\bar{C}_{\dot{\alpha}\dot{\beta}}(\bar{\p}^{\dot{\alpha}}
F)
(\bar{\p}^{\dot{\beta}} G)\nonumber\\
&&- \frac{1}{8}C^{\alpha\beta}C^{\gamma\delta}(\p_\alpha\p_\gamma F)\cdot(\p_\beta\p_\delta G)
- \frac{1}{8}\bar{C}_{\dot{\alpha}\dot{\beta}}\bar{C}_{\dot{\gamma}\dot{\delta}}
(\bar{\p}^{\dot{\alpha}}\bar{\p}^{\dot{\gamma}} F)(\bar{\p}^{\dot{\beta}}\bar{\p}^{\dot{\delta}} G) \nonumber\\
&&- \frac{1}{4}C^{\alpha\beta}\bar{C}_{\dot{\alpha}\dot{\beta}}
(\p_\alpha\bar{\p}^{\dot{\alpha}}F)(\p_\beta\bar{\p}^{\dot{\beta}} G) \nonumber\\
&& + \frac{1}{16}(-1)^{|F|}
C^{\alpha\beta}C^{\gamma\delta}\bar{C}_{\dot{\alpha}\dot{\beta}}
(\p_\alpha\p_\gamma\bar{\p}^{\dot{\alpha}} F)
(\p_\beta\p_\delta\bar{\p}^{\dot{\beta}} G) \nonumber\\
&& + \frac{1}{16}(-1)^{|F|}C^{\alpha\beta}\bar{C}_{\dot{\alpha}\dot{\beta}}
\bar{C}_{\dot{\gamma}\dot{\delta}}
(\p_\alpha\bar{\p}^{\dot{\alpha}}\bar{\p}^{\dot{\gamma}} F)(\p_\beta\bar{\p}^{\dot{\beta}}\bar{\p}^{\dot{\delta}} G)
\nonumber\\
&& + \frac{1}{64}C^{\alpha\beta}C^{\gamma\delta}
\bar{C}_{\dot{\alpha}\dot{\beta}}\bar{C}_{\dot{\gamma}\dot{\delta}}
(\p_\alpha\p_\gamma\bar{\p}^{\dot{\alpha}}\bar{\p}^{\dot{\gamma}}
F)(\p_\beta\p_\delta \bar{\p}^{\dot{\beta}}\bar{\p}^{\dot{\delta}}
G), \label{star}
\end{eqnarray}
where $|F| = 1$ if $F$ is odd (fermionic) and $|F|=0$ if $F$ is even (bosonic). In the
second line the definition of the multiplication $\mu_\star$ is given.
No higher powers of $C^{\alpha\beta}$ and
$\bar{C}_{\dot{\alpha}\dot{\beta}}$ appear since the derivatives
$\p_\alpha$ and $\bar{\p}^{\dot{\alpha}}$ are Grassmanian.
Expansion of the $\star$-product (\ref{star}) ends after the 4th
order in the deformation parameter. This is different from the
case of the Moyal-Weyl $\star_{\mbox{\tiny{mw}}}$-product
\cite{theta}, \cite{mw} where the expansion in powers of
the deformation parameter leads to an infinite power series. One should
also note that the $\star$-product (\ref{star}) is hermitian,
\begin{equation}
(F\star G)^* = G^* \star F^* , \label{complconj}
\end{equation}
where $*$ denotes the usual complex conjugation. This is
important for the construction of physical models.

The $\star$-product (\ref{star}) gives
\begin{eqnarray}
\{ \theta^\alpha \ds \theta^\beta \} &=& C^{\alpha\beta}, \quad \{
\bar{\theta}_{\dot\alpha}\ds \bar{\theta}_{\dot\beta}\} =
\bar{C}_{\dot{\alpha}\dot{\beta}}, \quad \{ \theta^\alpha \ds
\bar{\theta}_{\dot\alpha} \} = 0 ,\nonumber \\
\lbrack x^m \ds x^n \rbrack &=& 0 , \quad [x^m \ds \theta^\alpha ]
= 0, \quad [x^m \ds \bar{\theta}_{\dot\alpha} ] = 0 .
\label{thetastar}
\end{eqnarray}
Note that the chiral coordinates $y^m$ do not commute in this setting, but instead fulfill
\begin{eqnarray}
\lbrack y^m \ds y^n \rbrack &=&
-\theta\theta\bar{C}^{\dot{\alpha}\dot{\beta}}\varepsilon_{\dot{\beta}\dot{\gamma}}
(\bar{\sigma}^{mn})^{\dot{\gamma}}_{\ \dot{\alpha}}
-\bar{\theta}\bar{\theta}\varepsilon_{\alpha\beta}C^{\beta\gamma}(\sigma^{mn})_\gamma^{\ \alpha}, \nonumber\\
\lbrack y^m \ds \theta^\alpha \rbrack &=&
iC^{\alpha\beta}\sigma^m_{\
\beta\dot{\beta}}\bar{\theta}^{\dot{\beta}}, \quad \lbrack y^m \ds
\bar{\theta}_{\dot\alpha} \rbrack = i\theta^\alpha\sigma^m_{\
\alpha\dot{\beta}}\bar{C}^{\dot{\beta}\dot{\alpha}} .
\label{ystar}
\end{eqnarray}
Relations (\ref{thetastar}) enable us to define the deformed
superspace or "non\-anti\-commutative
space". It is generated by the usual bosonic and fermionic
coordinates (\ref{undefsupsp}) while the deformation is contained
in the new product (\ref{star}).

The deformed infinitesimal SUSY transformation is defined in the following way
\begin{eqnarray}
\delta^\star_\xi F &=& \big(\xi Q + \bar{\xi}\bar{Q} \big) F\nonumber\\
&=& X^\star_{\xi Q} \star F + X^\star_{\bar{\xi}\bar{Q}} \star F.
\label{defsusytr}
\end{eqnarray}
Differential operators $X^\star_{\xi Q}$ and $X^\star_{\bar{\xi}c}$
are given by
\begin{eqnarray}
X^\star_{\xi Q} &=& \xi^\alpha \Big( Q_\alpha +
\frac{1}{2}\bar{C}_{\dot{\beta}\dot{\gamma}}(\bar{\p}^{\dot{\beta}}
Q_\alpha)\bar{\p}^{\dot{\gamma}} \Big) \nonumber\\
&=& \xi^\alpha \Big( Q_\alpha +
\frac{i}{2}\bar{C}_{\dot{\beta}\dot{\gamma}}\sigma^m_{\
\alpha\dot{\alpha}}\varepsilon^{\dot{\alpha}\dot{\beta}}\p_m\bar{\p}^{\dot{\gamma}}
\Big) ,
\label{XQ}\\
X^\star_{\bar{\xi}\bar{Q}} &=& \bar{\xi}_{\dot{\alpha}} \Big(
\bar{Q}^{\dot{\alpha}} + \frac{1}{2}C^{\alpha\beta}(\p_\alpha
\bar{Q}^{\dot{\alpha}}) \p_\beta \Big) \nonumber\\
&=& \bar{\xi}_{\dot{\alpha}} \Big( \bar{Q}^{\dot{\alpha}} -
\frac{i}{2}C^{\alpha\beta}\sigma^m_{\ \alpha\dot{\gamma}}
\p_m\p_\beta \Big). \label{XbarQ}
\end{eqnarray}
Note that $X^\star$ operators close in the following algebra
\begin{equation}
\{ X^\star_{Q_\alpha} \ds X^\star_{Q_\beta} \} = \{
X^\star_{\bar{Q}^{\dot{\alpha}}} \ds
X^\star_{\bar{Q}^{\dot{\beta}}} \} = 0, \quad \{ X^\star_{Q_\alpha}
\ds X^\star_{\bar{Q}^{\dot{\beta}}} \} = 2i\sigma^m_{\
\alpha{\dot{\alpha}}}\p_m . \label{algX}
\end{equation}
This is just a different way of writing the algebra
(\ref{defQcom}). Differential operators $X^\star$ are mentioned
in \cite{Seiberg}, however no detailed analysis is preformed. In \cite{toppan} the authors discuss the Supersymmetric Quantum Mechanics with odd-parameters
being Clifford-valued and the operators similar to (\ref{XQ}) and (\ref{XbarQ}) arise.

The deformed coproduct (\ref{defcopr}) insures that the
$\star$-product of two superfields is again a superfield. Its
transformation law is given by
\begin{eqnarray}
\delta^\star_\xi (F\star G) &=& \big(\xi Q + \bar{\xi}\bar{Q} \big) (F\star G), \label{deftrlaw}\\
&=& \mu_\star\{ \Delta_{\cal{F}} (\delta^\star_\xi) F\otimes G) \}
,\nonumber
\end{eqnarray}
with
\begin{eqnarray}
\Delta_{\cal{F}} (\delta^\star_\xi) &=& {\cal F}
\Big( \delta^\star_\xi \otimes 1 + 1\otimes\delta^\star_\xi\Big){\cal F}^{-1}  \nonumber\\
&=& \delta^\star_\xi \otimes 1 + 1\otimes\delta^\star_\xi +
\frac{i}{2}C^{\alpha\beta}\Big( \bar{\xi}^{\dot{\gamma}}
\sigma^m_{\ \alpha\dot{\gamma}}\p_m\otimes\p_\beta +
\p_\beta\otimes \bar{\xi}^{\dot{\gamma}}
\sigma^m_{\ \alpha\dot{\gamma}}\p_m \Big) \nonumber\\
&& -\frac{i}{2}\bar{C}_{\dot{\alpha}\dot{\beta}} \Big(
\xi^\alpha\sigma^m_{\
\alpha\dot{\gamma}}\varepsilon^{\dot{\gamma}\dot{\alpha}}\p_m
\otimes \bar{\p}^{\dot{\beta}} + \bar{\p}^{\dot{\alpha}} \otimes
\xi^\alpha\sigma^m_{\ \alpha\dot{\gamma}}
\varepsilon^{\dot{\gamma}\dot{\beta}}\p_m \Big) .\nonumber
\end{eqnarray}
This gives
\begin{eqnarray}
\delta^\star_\xi (F\star G) &=&
(\delta^\star_\xi F) \star G + F \star (\delta^\star_\xi G) \nonumber\\
&& + \frac{i}{2}C^{\alpha\beta}\Big( \bar{\xi}^{\dot{\gamma}}
\sigma^m_{\ \alpha\dot{\gamma}}(\p_m F)\star (\p_\beta G) +
(\p_\alpha F)\star \bar{\xi}^{\dot{\gamma}}
\sigma^m_{\ \beta\dot{\gamma}}(\p_m G) \Big) \label{defLpravilo}\\
&& -\frac{i}{2}\bar{C}_{\dot{\alpha}\dot{\beta}} \Big(
\xi^\alpha\sigma^m_{\
\alpha\dot{\gamma}}\varepsilon^{\dot{\gamma}\dot{\alpha}}(\p_m F)
\star (\bar{\p}^{\dot{\beta}} G) + (\bar{\p}^{\dot{\alpha}} F)
\star \xi^\alpha\sigma^m_{\ \alpha\dot{\gamma}}
\varepsilon^{\dot{\gamma}\dot{\beta}}(\p_m G) \Big) .\nonumber
\end{eqnarray}

\section{Chiral fields}

Having established the general properties of the introduced deformation we
now turn to one special example, namely we study chiral fields. In
the undeformed theory chiral fields form a subalgebra of the
algebra of superfields. In the deformed case this will no longer
be the case.

A chiral field $\Phi$ fulfills $\bar{D}_{\dot{\alpha}}\Phi =0$,
where $\bar{D}_{\dot{\alpha}} = -\bar{\p}_{\dot{\alpha}} -
i\theta^\alpha \sigma^m_{\ \alpha\dot{\alpha}}\p_m$ is the
supercovariant derivative. In terms of component fields the chiral
superfield $\Phi$ is given by
\begin{eqnarray}
\Phi(x, \theta, \bar{\theta}) &=& A(x) + \sqrt{2}\theta^\alpha\psi_\alpha(x)
+ \theta\theta H(x) + i\theta\sigma^l\bar{\theta}(\p_l A(x)) \nonumber\\
&& -\frac{i}{\sqrt{2}}\theta\theta(\p_m\psi^\alpha(x))\sigma^m_{\ \alpha\dot{\alpha}}\bar{\theta}^{\dot{\alpha}}
+ \frac{1}{4}\theta\theta\bar{\theta}\bar{\theta}(\Box A(x)).\label{chiral}
\end{eqnarray}
Under the infinitesimal SUSY transformations (\ref{susytr}) component fields
transform as follows \cite{wessbook}
\begin{eqnarray}
\delta_\xi A &=& \sqrt{2}\xi\psi,\label{chtrA}\\
\delta_\xi \psi_\alpha &=& i\sqrt{2}\sigma^m_{\
\alpha\dot{\alpha}}\bar{\xi}^{\dot{\alpha}}(\p_m A)
+ \sqrt{2}\xi_\alpha H ,\label{chtrpsi}\\
\delta_\xi H &=& i\sqrt{2}\bar{\xi}\bar{\sigma}^m (\p_m \psi)
.\label{chtrF}
\end{eqnarray}

The $\star$-product of two chiral fields reads
\begin{eqnarray}
\Phi\star\Phi &=& A^2 - \frac{C^2}{2}H^2 +
\frac{1}{4}C^{\alpha\beta}\bar{C}^{\dot{\alpha}\dot{\beta}}
\sigma^m_{\ \alpha\dot{\alpha}}\sigma^l_{\ \beta\dot{\beta}}(\p_m
A)(\p_l A)
+\frac{1}{64}C^2\bar{C}^2 (\Box A)^2\nonumber\\
&& + \theta^\alpha\Big( 2\sqrt{2}\psi_\alpha A -\frac{1}{\sqrt{2}}C^{\gamma\beta}\bar{C}^{\dot{\alpha}\dot{\beta}}\varepsilon_{\gamma\alpha}
(\p_m\psi^\rho)\sigma^m_{\ \rho\dot{\beta}}\sigma^l_{\ \beta\dot{\alpha}}(\p_l A) \Big)
\nonumber\\
&& -\frac{i}{\sqrt{2}}C^2 \bar{\theta}_{\dot{\alpha}}
\bar{\sigma}^{m\dot{\alpha}\alpha}(\p_m\psi_\alpha)H
+ \theta\theta \Big( 2AH - \psi\psi\Big) \nonumber\\
&& + \bar{\theta}\bar{\theta}\Big( -\frac{C^2}{4}\big( H\Box A
- \frac{1}{2}(\p_m\psi)\sigma^m\bar{\sigma}^l(\p_l\psi) \big) \Big) \nonumber\\
&& + i\theta\sigma^m\bar{\theta}\Big( (\p_m A^2) + \frac{1}{4}C^{\alpha\beta}\bar{C}^{\dot{\alpha}\dot{\beta}}
\sigma_{m\alpha\dot{\alpha}}\sigma^l_{\ \beta\dot{\beta}}(\Box A)(\p_l A)\Big) \nonumber\\
&& +  i\sqrt{2}\theta\theta\bar{\theta}_{\dot{\alpha}}\bar{\sigma}^{m\dot{\alpha}\alpha}
\big( \p_m(\psi_\alpha A)\big) + \frac{1}{4}\theta\theta\bar{\theta}\bar{\theta} (\Box A^2) , \label{phistarphi}
\end{eqnarray}
where $C^2 =
C^{\alpha\beta}C^{\gamma\delta}\varepsilon_{\alpha\gamma}\varepsilon_{\beta\delta}$
and $\bar{C}^2 =
\bar{C}_{\dot{\alpha}\dot{\beta}}\bar{C}_{\dot{\gamma}\dot{\delta}}
\varepsilon^{\dot{\alpha}\dot{\gamma}}\varepsilon^{\dot{\beta}\dot{\delta}}$.
One sees that due to the $\bar{\theta}$ and the $\bar{\theta}\bar{\theta}$ terms (\ref{phistarphi}) is not a chiral field. However, in order to write an action invariant under the deformed
SUSY transformations (\ref{defsusytr}) we need to preserve the
notion of chirality. This can be done in different ways. One possibility is to use a different $\star$-product, the one which preserves
chirality \cite{chiralstar}. However, chirality-preserving
$\star$-product implies working in Euclidean space where
$\bar{\theta} \neq (\theta)^*$. Since we want to work in Minkowski
space-time we use the
$\star$-product (\ref{star}) and decompose $\star$-products of superfields
into their irreducible components by using the projectors defined in
\cite{wessbook}.

The chiral, antichiral and transversal projectors are defined as
follows
\begin{eqnarray}
P_1 &=& \frac{1}{16} \frac{D^2 \bar{D}^2}{\Box}, \label{P1}\\
P_2 &=& \frac{1}{16} \frac{\bar{D}^2 D^2 }{\Box}, \label{P2}\\
P_T &=& -\frac{1}{8} \frac{D \bar{D}^2 D}{\Box}. \label{PT}
\end{eqnarray}
In order to calculate irreducible components of the $\star$-products
of chiral superfields, we first apply the projectors
(\ref{P1})-(\ref{PT}) to the superfield $F$ (\ref{F}). From the
definition of the supercovariant derivatives
\begin{eqnarray}
D_\alpha &=& \p_\alpha
+ i\sigma^m_{\ \alpha\dot{\alpha}}\bar{\theta}^{\dot{\alpha}}\p_m, \label{D}\\
\bar{D}_{\dot{\alpha}} &=& -\bar{\p}_{\dot{\alpha}} -
i\theta^\alpha \sigma^m_{\ \alpha\dot{\alpha}}\p_m, \label{barD}
\end{eqnarray}
follows
\begin{eqnarray}
D^2 &=& D^\alpha D_\alpha =
-\varepsilon^{\alpha\beta}\p_\alpha\p_\beta +
2i\varepsilon^{\alpha\beta}
\sigma^m_{\ \beta\dot{\beta}}\bar{\theta}^{\dot{\beta}}\p_\alpha\p_m - \bar{\theta}\bar{\theta} \Box ,\label{D^2}\\
\bar{D}^2 &=& \bar{D}_{\dot{\alpha}}\bar{D}^{\dot{\alpha}} =
\varepsilon^{\dot{\alpha}\dot{\beta}}\bar{\p}_{\dot{\alpha}}{\bar\p}_{\dot{\beta}}
+ 2i\theta^\alpha\sigma^m_{\ \alpha\dot{\alpha}}
\varepsilon^{\dot{\alpha}{\dot{\beta}}} {\bar\p}_{\dot{\beta}}\p_m
- \theta\theta \Box .\label{barD^2}
\end{eqnarray}

Let us start with $P_2$ and calculate first
\begin{eqnarray}
D^2F &=& -4m -2 \bar{\theta}_{\dot{\alpha}}\Big(
2\bar{\lambda}^{\dot{\alpha}} +
i\bar{\sigma}^{m\dot{\alpha}\alpha}(\p_m\phi_\alpha) \Big)
+4i\theta\sigma^l\bar{\theta}(\p_l m) \nonumber\\
&& - \bar{\theta}\bar{\theta}\Big( 4d + \Box f -2i (\p_m v^m)\Big) \nonumber\\
&& - \bar{\theta}\bar{\theta}\theta^\alpha \Big( 2i\sigma^m_{\ \alpha\dot{\alpha}} (\p_m \bar{\lambda}^{\dot{\alpha}}) + (\Box\phi_\alpha) \Big)
- \theta\theta\bar{\theta}\bar{\theta}(\Box m) .\label{D^2F}
\end{eqnarray}
Then we have
\begin{eqnarray}
\bar{D}^2D^2 F &=& 4\Big( 4d + \Box f -2i (\p_m v^m)\Big) +
8\theta^\alpha \Big( 2i\sigma^m_{\ \alpha\dot{\alpha}}
(\p_m \bar{\lambda}^{\dot{\alpha}}) + (\Box\phi_\alpha) \Big) \nonumber\\
&& + 16\theta\theta (\Box m) + 4i\theta\sigma^l\bar{\theta}
\Big( 4\p_l d + \p_l\Box f -2i (\p_m\p_l v^m)\Big)\nonumber\\
&& + 4\theta\theta\bar{\theta}_{\dot{\alpha}} \Big( 2\Box\bar{\lambda}^{\dot{\alpha}}
+ i\bar{\sigma}^{m\dot{\alpha}\alpha}(\p_m\Box\phi_\alpha) \Big) \nonumber\\
&& + \theta\theta\bar{\theta}\bar{\theta} \Big( 4\Box d + \Box^2 f -2i\Box\p_m v^m \Big) .\label{barD^2D^2F}
\end{eqnarray}
This gives
\begin{eqnarray}
P_2F&=& \frac{1}{16} \frac{\bar{D}^2 D^2 }{\Box} F \nonumber\\
&=& \frac{1}{\Box}\Big( d - \frac{i}{2}(\p_m v^m) +
\frac{1}{4}\Box f\Big) + \sqrt{2}\theta^\alpha\Big(
\frac{i}{\sqrt{2}\Box}\sigma^m_{\ \alpha\dot{\alpha}}
(\p_m \bar{\lambda}^{\dot{\alpha}}) + \frac{1}{2\sqrt{2}}\phi_\alpha \Big) \nonumber\\
&& + \theta\theta m
+ i\theta\sigma^l \bar{\theta}\p_l\Big( \frac{d}{\Box} - \frac{i}{2\Box}(\p_m v^m)
+ \frac{1}{4}f \Big) \label{P2F'}\\
&& + \frac{1}{\sqrt{2}}\theta\theta\bar{\theta}_{\dot\alpha}\Big( \frac{1}{\sqrt{2}}\bar{\lambda}^{\dot{\alpha}}
+ \frac{i}{2\sqrt{2}}\bar{\sigma}^{m\dot{\alpha}\alpha}(\p_m\phi_\alpha) \Big)
+\frac{1}{4}\theta\theta\bar{\theta}\bar{\theta}\Big(
d - \frac{i}{2}(\p_m v^m) + \frac{1}{4}\Box f \Big) .\nonumber
\end{eqnarray}
The superfield (\ref{P2F'}) is a chiral field with the components
\begin{eqnarray}
&&{\mbox{scalar: }}\quad {\cal A} = \frac{1}{\Box}\Big( d - \frac{i}{2}(\p_m v^m)
+ \frac{1}{4}\Box f\Big) ,\label{Aizphistarphi}\\
&&{\mbox{spinor: }}\quad {\cal \psi}_\alpha = \frac{i}{\sqrt{2}\Box}\sigma^m_{\ \alpha\dot{\alpha}}
(\p_m \bar{\lambda}^{\dot{\alpha}}) + \frac{1}{2\sqrt{2}}\phi_\alpha, \label{psiizphistarphi}\\
&&{\mbox{auxiliary field: }} \quad {\cal H} = m. \label{Fizphistarphi}
\end{eqnarray}
In general, some of these component fields will be nonlocal due
to $1/\Box$ in the definition of the projector $P_2$.

A calculation analogous to the previous one leads to
\begin{eqnarray}
P_1F&=& \frac{1}{16} \frac{D^2\bar{D}^2}{\Box} F \nonumber\\
&=& \frac{1}{\Box}\Big( d + \frac{i}{2}(\p_m v^m) +
\frac{1}{4}\Box f\Big) + \sqrt{2}\bar{\theta}_{\dot\alpha}\Big(
\frac{i}{\sqrt{2}\Box}\bar{\sigma}^{m\dot{\alpha}\alpha}
(\p_m \varphi_\alpha) + \frac{1}{2\sqrt{2}}\bar{\chi}^{\dot\alpha} \Big) \nonumber\\
&& + \bar{\theta}\bar{\theta} n
- i\theta\sigma^l \bar{\theta}\p_l\Big( \frac{d}{\Box} + \frac{i}{2\Box}(\p_m v^m)
+ \frac{1}{4}f \Big) \label{P1F'}\\
&& - \frac{1}{\sqrt{2}}\bar{\theta}\bar{\theta}\theta^\alpha\Big( \frac{1}{\sqrt{2}}\varphi_\alpha
- \frac{i}{2\sqrt{2}}\sigma^m_{\ \alpha\dot{\alpha}}(\p_m\bar{\chi}^{\dot\alpha}) \Big)
+\frac{1}{4}\theta\theta\bar{\theta}\bar{\theta}\Big(
d + \frac{i}{2}(\p_m v^m) + \frac{1}{4}\Box f \Big) ,\nonumber
\end{eqnarray}
which is an antichiral field with the components
\begin{eqnarray}
&&{\mbox{scalar: }}\quad \widetilde{{\cal A}} = \frac{1}{\Box}\Big(
d + \frac{i}{2}(\p_m v^m)
+ \frac{1}{4}\Box f\Big) ,\label{AizP1F}\\
&&{\mbox{spinor: }}\quad \overline{\widetilde{{\cal
\psi}}}^{\dot{\alpha}} =
\frac{i}{\sqrt{2}\Box}\bar{\sigma}^{m\dot{\alpha}\alpha}
(\p_m \varphi_\alpha) + \frac{1}{2\sqrt{2}}\bar{\chi}^{\dot\alpha}, \label{psiizP1F}\\
&&{\mbox{auxiliary field: }} \quad \widetilde{{\cal H}} = n.
\label{FizP1F}
\end{eqnarray}

For the completeness we give the action of the transversal
projector $P_T$ on the superfield (\ref{F}). It follows from the identity
\begin{equation}
P_T = I-P_1-P_2 .\label{PT'}
\end{equation}
By using (\ref{P2F'}) and (\ref{P1F'}) we obtain
\begin{eqnarray}
P_T F &=& \frac{1}{2}f - \frac{2}{\Box}d +\theta^\alpha\Big(
\frac{1}{2}\phi_\alpha-
i\frac{1}{\Box}\sigma^m_{\ \alpha\dot{\alpha}}\p_m\bar{\lambda}^{\dot\alpha}\Big) \nonumber\\
&& +\bar\theta_{\dot{\alpha}}\Big( \frac{1}{2}\bar\chi^{\dot{\alpha}}-i\frac{1}{\Box}\bar{\sigma}^{m\dot{\alpha}\alpha}
\p_m\varphi_{\alpha} \Big)
+\theta\sigma^m\bar{\theta} \Big( v_m-\frac{1}{\Box}\p_m\p_l v^l \Big)\nonumber\\
&& +\theta\theta\bar{\theta}_{\dot{\alpha}}\Big( \frac{1}{2}
\bar{\lambda}^{\dot\alpha} - \frac{i}{4}\bar{\sigma}^{m\dot{\alpha}\alpha}(\p_m\phi_\alpha)\Big)
+ \bar{\theta}\bar{\theta}\theta^\alpha\Big( \frac{1}{2}
\varphi_\alpha - \frac{i}{4}\sigma^m_{\ \alpha\dot{\alpha}}(\p_m\bar{\chi}^{\dot\alpha})\Big)\nonumber\\
&& +\frac{1}{4}\theta\theta\bar{\theta}\bar{\theta} \Big( 2d - \frac{1}{2}\Box f\Big) .\label{PTF}
\end{eqnarray}

\section{Deformed Wess-Zumino Lagrangian}

In the undeformed theory, Wess-Zumino Lagrangian is given by
\begin{eqnarray}
{\cal L} = \Phi^+\cdot\Phi\Big|_{\theta\theta\bar\theta\bar\theta}
+ \Big( \frac{m}{2} \Phi\cdot\Phi \Big|_{\theta\theta} +
\frac{\lambda}{3}\Phi\cdot\Phi\cdot\Phi \Big |_{\theta\theta} +
{\mbox{ c.c. }} \Big) , \label{undefL}
\end{eqnarray}
where $m$ and $\lambda$ are real constants, $\Phi$ is a chiral field
and $\Phi^+$ is an antichiral field with $(\Phi^+)^+ =\Phi$. This
Lagrangian leads to the SUSY invariant action which describes
an interacting theory of two complex scalar fields and one spinor
field. To see this explicitly we look at each term separately.
This analysis is well known but we repeat it nevertheless to
prepare for the analysis of the deformed Wess-Zumino
Lagrangian.

The kinetic term is given by the highest component of the product
$\Phi^+\cdot \Phi$:
\begin{equation}
\Phi^+\cdot \Phi\Big|_{\theta\theta\bar{\theta}\bar{\theta}} =
A^*\Box A + i(\p_m\bar\psi)\bar{\sigma}^m\psi +
H^*H. \label{undefkin}
\end{equation}
Since $\Phi^+\cdot \Phi$ is a superfield, its highest component
has to transform as a total derivative, (\ref{susytrd}).

Next we look at the mass term. It is given by the $\theta\theta$
component of $\Phi\cdot\Phi$ and the $\bar{\theta}\bar{\theta}$
component of $\Phi^+\cdot\Phi^+$:
\begin{equation}
\frac{m}{2}\Big( \Phi\cdot\Phi\Big|_{\theta\theta} +
\Phi^+\cdot\Phi^+\Big|_{\bar{\theta}\bar{\theta}}\Big) =
\frac{m}{2}\Big( 2AH - \psi\psi + 2A^*H^* -
\bar{\psi}\bar{\psi} \Big) .\label{undefmass}
\end{equation}
As the pointwise product of two chiral/antichiral fields is a
chiral/antichiral field, its
$\theta\theta$/$\bar{\theta}\bar{\theta}$ component transforms as
a total derivative (\ref{chtrF}). Note that this is not the case
with the general superfield (\ref{susytrm}). Also note that the
highest components of $\Phi\cdot\Phi$ and $\Phi^+\cdot\Phi^+$
transform as total derivatives. However, these terms are total
derivatives themselves (\ref{chiral}) and will not contribute to
the equations of motion.

The same arguments apply for the interaction term, since
$\Phi\cdot\Phi\cdot\Phi$ is a chiral field again and
$\Phi^+\cdot\Phi^+\cdot\Phi^+$ is an antichiral field. The
interaction term reads
\begin{equation}
\frac{\lambda}{3}\Big( \Phi\cdot\Phi\cdot\Phi\Big|_{\theta\theta}
+
\Phi^+\cdot\Phi^+\cdot\Phi^+\Big|_{\bar{\theta}\bar{\theta}}\Big)
= \frac{\lambda}{3}\Big( HA^2 - A\psi\psi +
H^*(A^*)^2 - A^*\bar\psi\bar\psi \Big) .\label{undefint}
\end{equation}
Thus, we see that chirality plays an important role in
the construction of a SUSY invariant action.

We are interested in a deformation of (\ref{undefL}) which is
consistent with the deformed SUSY transformations
(\ref{defsusytr}) and which in the limit $C^{\alpha\beta}\to 0$ gives the undeformed Lagrangian (\ref{undefL}).

We propose the following Lagrangian
\begin{eqnarray}
{\cal L} = \Phi^+\star\Phi\Big|_{\theta\theta\bar\theta\bar\theta}
+ \Big( \frac{m}{2}P_2\big( \Phi\star\Phi\big)
\Big|_{\theta\theta} + \frac{\lambda}{3}P_2\Big( \Phi\star
P_2\big( \Phi\star\Phi\big)\Big) \Big |_{\theta\theta} + {\mbox{
c.c }} \Big) , \label{L}
\end{eqnarray}
where $m$ and $\lambda$ are real constants. Let us analyse
(\ref{L}) term by term again.

Kinetic term in (\ref{L}) is a straightforward deformation of the
usual kinetic term obtained by inserting the $\star$-product instead the
usual pointwise multiplication. Due to the deformed coproduct
(\ref{defcopr}), $\Phi^+ \star\Phi$ is a superfield and its highest
component transforms as a total derivative. The explicit calculation
gives
\begin{eqnarray}
\Phi^+ \star\Phi\Big|_{\theta\theta\bar{\theta}\bar{\theta}} &=&
A^*\Box A
+ i(\p_m\bar\psi)\bar{\sigma}^m\psi + H^*H , \label{defkin} \\
\delta^\star_\xi \Big( \Phi^+
\star\Phi\Big|_{\theta\theta\bar{\theta}\bar{\theta}} \Big) &=&
\p_m\Big( \frac{1}{2\sqrt{2}} \big( A^*(\p_l\psi^\alpha) - (\p_l
A^*)\psi^\alpha\big) (\sigma^l\bar{\sigma}^m)_\alpha^{\ \beta} +
\frac{i}{\sqrt{2}}H\bar{\psi}_{\dot{\alpha}}
\bar{\sigma}^{m\dot{\alpha}\beta} \Big) \xi_\beta \nonumber\\
&& +\bar{\xi}_{\dot{\alpha}} \p_m \Big( \frac{1}{2\sqrt{2}}
(\bar{\sigma}^m \sigma^l)^{\dot{\alpha}}_{\ \dot{\beta}}
\big( \bar{\psi}^{\dot{\beta}}(\p_l A) - (\p_l\bar{\psi}^{\dot{\beta}})A \big)
+ \frac{i}{\sqrt{2}}\bar{\sigma}^{m\dot{\alpha}\alpha}H^*\psi_\alpha \Big)
.\label{trlawdefkin}
\end{eqnarray}
To obtain (\ref{defkin}), the partial integration was used. We see
from (\ref{defkin}) that the deformation is absent,
the kinetic term remains undeformed\footnote{ In the case of the
Moyal-Weyl $\star$-product we have $\int{\mbox{d}}^4 x\hspace{1mm}
f\star_{{\mbox{\tiny{mw}}}} g = \int{\mbox{d}}^4 x\hspace{1mm}
g\star_{{\mbox{\tiny{mw}}}} f = \int{\mbox{d}}^4 x\hspace{1mm}
f\cdot g$. Therefore, the free actions for scalar and spinor
fields remain undeformed automatically.}.

Since $\Phi\star\Phi$ is not a chiral field we have to
project its chiral part. This projection is given by
\begin{eqnarray}
P_2\big( \Phi\star\Phi \big) &=& A^2 -
 \frac{C^2}{8}H^2 + \frac{1}{256}C^2\bar{C}^2 (\Box A)^2\nonumber\\
&& + \frac{1}{16}C^{\alpha\beta}\bar{C}^{\dot{\alpha}\dot{\beta}}
\sigma^m_{\ \alpha\dot{\alpha}}\sigma^l_{\ \beta\dot{\beta}} \Big( (\p_m A)(\p_l A)
+ \frac{2}{\Box}\p_m \big( (\Box A)(\p_l A) \big) \Big) \nonumber\\
&& + \sqrt{2}\theta^\alpha \Big( 2\psi_\alpha A -\frac{1}{4}
C^{\gamma\beta}\bar{C}^{\dot{\alpha}\dot{\beta}}\varepsilon_{\gamma\alpha}
(\p_m\psi^\rho)\sigma^m_{\ \rho\dot{\beta}}\sigma^l_{\
\beta\dot{\alpha}}(\p_l A) \Big)
\nonumber\\
&& + \theta\theta \Big( 2AH - \psi\psi\Big) \nonumber\\
&& + i\theta\sigma^k\bar{\theta}\p_k \Big[ A^2 - \frac{C^2}{8}H^2 + \frac{1}{256}C^2\bar{C}^2 (\Box A)^2\nonumber\\
&&+ \frac{1}{16}C^{\alpha\beta}\bar{C}^{\dot{\alpha}\dot{\beta}}
\sigma^m_{\ \alpha\dot{\alpha}}\sigma^l_{\ \beta\dot{\beta}} \Big(
(\p_m A)(\p_l A) + \frac{2}{\Box}\p_m \big( (\Box A)(\p_l A) \big)
\Big)
\Big]\nonumber\\
&& + i\sqrt{2} \theta\theta\bar{\theta}_{\dot{\alpha}}
\bar{\sigma}^{k\dot{\alpha}\alpha}\p_k \Big( \psi_\alpha A
-\frac{1}{8}
C^{\gamma\beta}\bar{C}^{\dot{\alpha}\dot{\beta}}\varepsilon_{\gamma\alpha}
(\p_m\psi^\rho)\sigma^m_{\ \rho\dot{\beta}}\sigma^l_{\ \beta\dot{\alpha}}(\p_l A) \Big) \nonumber\\
&&+ \frac{1}{4}\theta\theta\bar{\theta}\bar{\theta}\Box
\Big[ A^2 - \frac{C^2}{8}H^2 + \frac{1}{256}C^2\bar{C}^2 (\Box A)^2\nonumber\\
&& + \frac{1}{16}C^{\alpha\beta}\bar{C}^{\dot{\alpha}\dot{\beta}}
\sigma^m_{\ \alpha\dot{\alpha}}\sigma^l_{\ \beta\dot{\beta}} \Big(
(\p_m A)(\p_l A) + \frac{2}{\Box}\p_m \big( (\Box A)(\p_l A) \big)
\Big) \Big] .\label{P2phistarphi}
\end{eqnarray}
For the action we take the $\theta\theta$ component of
(\ref{P2phistarphi}),
\begin{equation}
P_2\big( \Phi\star\Phi \big)\Big|_{\theta\theta} = 2AH -
\psi\psi .\label{masstr}
\end{equation}
Its transformation law is given by
\begin{equation}
\delta^\star_\xi\Big( P_2\big( \Phi\star\Phi
\big)\Big|_{\theta\theta} \Big)  = 2i\sqrt2\bar{\xi}\bar{\sigma}^m
\p_m(A\psi) .\label{deltamasstr}
\end{equation}
In a similar way we add the $\bar{\theta}\bar{\theta}$
component of $P_1(\Phi^+\star \Phi^+)$. This component is given by
\begin{equation}
P_1\big( \Phi^+\star\Phi^+ \big)\Big|_{\bar{\theta}\bar{\theta}} =
2A^*H - \bar{\psi}\bar{\psi} ,\label{masstr2}
\end{equation}
which is just the complex conjugate of (\ref{masstr}) due to the
hermiticity of the $\star$-product
(\ref{star}). Again, no deformation is present: the free action remains undeformed. That leads to the propagators which are the same as in the undeformed theory.

Finally we come to the interaction term. There are few possibilities to project the chiral part of $\Phi\star\Phi\star\Phi$. We
take the following projection\footnote{Naively, one would take
$P_2(\Phi\star\Phi\star\Phi)\big|_{\theta\theta}$. Despite the
fact that $P_2(\Phi\star\Phi\star\Phi)$ is a chiral field, its
$\theta\theta$ component does not transform as a total derivative
and would not lead to a SUSY invariant action. This strange
situation arises because of the $1/\Box$ term in the projector $P_2$.}
\begin{equation}
\Phi\star\Phi\star\Phi \to P_2\Big( \Phi\star \big( P_2
(\Phi\star\Phi)\big)\Big). \label{defint}
\end{equation}
As the complete result is very long we write here only the
$\theta\theta$ component
\begin{eqnarray}
P_2\Big( \Phi\star \big( P_2
(\Phi\star\Phi)\big)\Big)\Big|_{\theta\theta}&=& 3\big( A^2H -
(\psi\psi)A\big) -\frac{C^2}{8}H^3
+ \frac{1}{256}C^2\bar{C}^2H(\Box A)^2 \nonumber\\
&& +\frac{1}{16}C^{\al\be}\bar
C^{\dot{\al}\dot{\be}}\si_{\al\dot{\al}}^m\si^l_{\be\dot{\be}}
H\Big( (\p_m A)(\p_l A) + \frac{2}{\Box}\p_m \big( (\Box A)(\p_l A) \big) \Big)\nonumber\\
&&+\frac{1}{4}C^{\gamma\be}\bar
C^{\dot\al\dot\be}H\si^l_{\be\dot\al}
\psi_\gamma(\p_m\psi^\rho)\si^m_{\rho\dot\be}
(\p_l A) \label{FizP2phiP2phiphi}\\
&&+
\frac{1}{2}\bar{C}_{\dot{\alpha}\dot{\beta}}(\bar{\sigma}^{lm})^{\dot{\beta}}_{\
\dot{\gamma}} \varepsilon^{\dot{\gamma}\dot{\alpha}}(\p_m
A)\p_l\Big[
A^2 -\frac{C^2}{8}H^2 \nonumber\\
&& +\frac{1}{16}C^{\al\be}\bar C^{\dot\al\dot\be}\si^s_{\
\al\dot\al}\si^p_{\ \be\dot\be}H\Big( (\p_s A)(\p_p A) +
\frac{2}{\Box}\p_s \big( (\Box A)(\p_p A) \big) \Big) \Big] .\nn
\end{eqnarray}
In the limit $C^{\alpha\beta}\to 0$ (\ref{FizP2phiP2phiphi}) reduces to the
usual interaction term (\ref{undefint}). The deformation is
present trough the terms that are of first, second and higher orders in
$C^{\alpha\beta}$ and $\bar{C}_{\dot{\alpha}\dot{\beta}}$. Note that under the integral the last term reduces to a total
derivative and therefore will not contribute to the equations of
motion. Also note that if we calculate $P_2\Big( \big( P_2
(\Phi\star\Phi)\big)\star\Phi\Big)$ instead of (\ref{defint}) the
only difference will be in the sign of the above-mentioned last term. We therefore conclude that we can
take any combination of these two terms, as long as the limit $C^{\alpha\beta}
\to 0$ reproduces the undeformed interaction term. For simplicity
we take only (\ref{FizP2phiP2phiphi}).

The transformation law of (\ref{FizP2phiP2phiphi}) is given by
\begin{eqnarray}
&&\delta^\star_\xi\Big( P_2\Big( \Phi\star \big( P_2
(\Phi\star\Phi)\big)\Big)\Big|_{\theta\theta}\Big)= \nonumber\\
&&i\sqrt{2}\bar\xi_{\dot{\al}}\bar\si^{l\dot{\al}\al}\p_l\Big(
\frac{1}{8}C^{\gamma\beta}\bar{C}^{\dot{\gamma}\dot{\beta}} \sigma^m_{\
\gamma\dot{\gamma}}\sigma^n_{\ \beta\dot{\beta}} \psi_{\al}\frac{1}{\Box}\p_m(\p_n A\Box A) + {\mbox{ local
terms}}\Big) . \label{trlawintterm}
\end{eqnarray}
The SUSY transformation is a total derivative
and reduces to a surface term under the integral, leading to a
SUSY invariant interaction term. However, one should be careful
as (\ref{trlawintterm}) contains a non-local term. Under the
integral it is proportional to
\begin{eqnarray}
&&\int{\mbox{d}}^4 x\hspace{1mm}\bar\si^{l\dot{\al}\al}\p_l\Big(
\psi_{\al}\frac{1}{\Box}\p_m(\p_n A\Box A) \Big) \nonumber\\
&=& \oint{\mbox{d}}\Sigma_l\hspace{1mm}
\bar\si^{l\dot{\al}\al}\Big( \psi_{\al}\frac{1}{\Box}\p_m(\p_n
A\Box A) \Big). \nonumber
\end{eqnarray}
If the boundary surface $\Sigma_l$ is at infinity and fields
fall off fast enough this integral vanishes.

To rewrite (\ref{FizP2phiP2phiphi}) in a more compact way we
introduce the following notation
\begin{eqnarray}
C_{\alpha\beta}&=& K_{ab}(\sigma^{ab}\varepsilon)_{\alpha\beta},\label{n1}\\
\bar{C}_{\dot{\alpha}\dot{\beta}}&=&
K^*_{ab}(\varepsilon\bar{\sigma}^{ab})_{\dot{\alpha}\dot{\beta}},\label{n2}
\end{eqnarray}
where $K_{ab}=-K_{ba}$ is an antisymmetric complex
constant matrix. Then we have
\begin{eqnarray}
C^2 = 2K_{ab}K^{ab}, && \bar{C}^2 = 2K^*_{ab}K^{*ab}, \quad
K^{ab}K^*_{ab}= 0.\label{n4} \\
K^*_{cd}K_{ab}\big( \sigma^n\bar{\sigma}^{cd}\bar{\sigma}^m
\sigma^{ab}\big)_\alpha^{\ \beta} &=&
-4\delta^\beta_\alpha K^{ma}K^{*n}_{\ \ a} + 8 K^{ma}K^{*nb}(\sigma_{ba})_\alpha^{\ \beta} ,\label{n5}\\
C^{\alpha\beta}\bar{C}^{\dot{\alpha}\dot{\beta}} \sigma^m_{\
\alpha\dot{\alpha}}\sigma^l_{\ \beta\dot{\beta}} &=& 8K^{am}K^{*\
l}_a. \label{n6}
\end{eqnarray}

By using the previous expressions the term (\ref{FizP2phiP2phiphi}) can be
rewritten in the form
\begin{eqnarray}
P_2\Big( \Phi\star \big( P_2
(\Phi\star\Phi)\big)\Big)\Big|_{\theta\theta}&=& 3\big( A^2H -
(\psi\psi)A\big)
-\frac{1}{4}\KKt H^3 \nonumber\\
&&+ \frac{1}{64}\KKt\KKtc H(\Box A)^2 \label{FizP2phiP2phiphiK}\\
&& +\frac{1}{2}\KK H\Big( (\p_m A)(\p_n A) + \frac{2}{\Box}\p_m \big( (\Box A)(\p_n A) \big) \Big)\nonumber\\
&&-\Big(\KK\psi(\p_n\psi) - 2K^m_{\ a}K^{*n}_{\ \
c}(\p_n\psi)\si^{ca}\psi\Big)(\p_m A) .\nonumber
\end{eqnarray}

Finally, the deformed SUSY invariant Lagrangian is given by
\begin{eqnarray}
{\cal L} &=&
\Phi^+\star\Phi\Big|_{\theta\theta\bar\theta\bar\theta}\nonumber\\
&& + \Big( \frac{m}{2}P_2\big( \Phi\star\Phi\big)
\Big|_{\theta\theta} + \frac{\lambda}{3}P_2\Big( \Phi\star P_2\big(
\Phi\star\Phi\big)\Big) \Big |_{\theta\theta} + {\mbox{ c.c }} \Big) \nonumber\\
&=&
A^*\Box A + i(\p_m\bar\psi)\bar{\sigma}^m\psi + H^*H\nonumber\\
&& + \frac{m}{2}\Big( 2AH - \psi\psi + 2A^*H^* - \bar{\psi}\bar{\psi} \Big) \nonumber\\
&&
+\lambda\Big( HA^2 - A\psi\psi + H^*(A^*)^2 - A^*\bar\psi\bar\psi \Big)\nonumber\\
&& -\frac{\lambda}{3}\Big(K^{m}_{\ a}K^{*na}\psi(\p_n\psi) - 2K^m_{\
a}K^{*n}_{\ \ b}(\p_n\psi)\sigma^{ba}\psi\Big) (\p_m A)\nonumber\\
&& -\frac{\lambda}{3}\Big(K^{m}_{\
a}K^{*na}\bar\psi(\p_n\bar\psi) - 2K^{*m}_{\ \ a}K^{n}_{ \
b}\bar\psi\bar\sigma^{ab}(\p_n\bar\psi) \Big) (\p_mA^*) \nonumber\\
&& -\frac{\lambda}{12}K^{mn}K_{mn}H^3
-\frac{\lambda}{12}K^{*mn}K_{mn}^*( H^*)^3\nonumber\\
&&+\frac{\la}{6}\KK\Big( H(\p_m A)(\p_n A)
+ H^*(\p_m A^*)(\p_n A^*) \Big)\nn\\
&& +\frac{\la}{3}\KK\Big[ H\frac{1}{\Box}\p_m\Big( (\p_n A)\Box
A\Big) + H^*\frac{1}{\Box}\p_m\Big( (\p_n A^*)\Box A^*\Big)\Big]\nn\\
&&+\frac{\la}{192}\KKt\KKtc \Big( H(\Box A)^2 + H^*(\Box A)^*\Big),
\label{Lincompfields}
\end{eqnarray}
where the partial integration was used to rewrite some of the terms in
(\ref{Lincompfields}) in a more compact way.

\section{Equations of motion}

By varying the action which follows from the Lagrangian
(\ref{Lincompfields}) with respect to the fields $H$ and
$H^*$ we obtain the equations of motion
\begin{eqnarray}
H^* &+& mA + \lambda A^2 -\frac{\la}{4}\KKt H^2
+ \frac{\la}{6}\KK(\p_m A)(\p_n A) \nn\\
&+&\frac{\la}{3}\KK\frac{1}{\Box}\p_m\big( (\p_n A)\Box A\big) + \frac{\la}{192}\KKt\KKtc(\Box A)^2= 0 ,\label{eomF*}\\
H &+& mA^* + \lambda(A^*)^2 -\frac{\la}{4}\KKtc(H^*)^2+\frac{\la}{6}\KK(\p_m A^*)(\p_n A^*)\nn\\
&+&\frac{\la}{3}\KK\frac{1}{\Box}\p_m\big( (\p_n A^*)\Box A^*\big)
+ \frac{\la}{192}\KKt\KKtc(\Box A^*)^2= 0 . \label{eomF}
\end{eqnarray}
Unlike the undeformed theory, equations (\ref{eomF*}) and
(\ref{eomF}) are nonlinear in $H$ and $H^*$.
Nevertheless, they can be solved perturbatively. The solutions are given by
\begin{eqnarray}
H^* &=& -mA -\la A^2+\frac{\la}{4}\KKt (mA^*+\la (A^*)^2)^2\nn\\
&& -\frac{\la}{6}\KK(\p_m A)(\p_n A) - \frac{\la}{3}\KK\frac{1}{\Box}\p_m\big( (\p_n A)\Box A\big) \nn\\
&& -\frac{\la}{192}\KKt\KKtc(\Box A)^2
\nonumber\\
&&+\frac{\la}{2}\KKt(mA^*
+\la(A^*)^2)\Big[\frac{\la}{6}\KK (\p_m
A^*)(\p_n A^*) \label{pertsolF}\\
&& + \frac{\la}{3}\KK\frac{1}{\Box}\p_m\big( (\p_n A^*)\Box
A^*\big)
+\frac{\la}{4}\KKtc(mA+\la A^2)^2\Big] + {\cal O}(K^6) ,\nonumber\\
H &=& -mA^* -\la (A^*)^2+\frac{\la}{4}\KKtc (mA+\la A^2)^2 -\frac{\la}{6}\KK(\p_m A^*)(\p_n A^*)\nn\\
&& - \frac{\la}{3}\KK\frac{1}{\Box}\p_m\big( (\p_n A^*)\Box A^*\big) -\frac{\la}{192}\KKt\KKtc(\Box A^*)^2\nn\\
&& +\frac{\la}{2}\KKtc(mA+\la A^2)\Big[ \frac{\la}{6}\KK (\p_m
A)(\p_n A)\label{pertsolF*}\\
&&+ \frac{\la}{3}\KK\frac{1}{\Box}\p_m\big( (\p_n A)\Box
A)+\frac{\la}{4}\KKt(mA^*+\la (A^*)^2)^2\Big] +{\cal O}(K^6)
.\nonumber
\end{eqnarray}
These solutions can be used to
eliminate the auxiliary fields $H$ and $H^*$
from the Lagrangian (\ref{Lincompfields}). This gives
\begin{equation}
{\cal L}={\cal L}_0+{\cal L}_2+{\cal L}_4+{\cal O}(K^6)\ ,
\label{LohneF}
\end{equation}
with
\begin{eqnarray}
{\cal L}_0 &=& A^*\Box A + i(\p_m\bar\psi)\bar{\sigma}^m\psi
-\lambda A^*\bar\psi\bar\psi - \lambda A\psi\psi -
\frac{m}{2}\big( \psi\psi
+ \bar\psi\bar\psi \big) \nonumber\\
&& -m^2A^*A-m\la A(A^*)^2-m\la A^*A^2-\la^2A^2(A^*)^2\
,\label{L0bezF} \\
{\cal L}_2 &=&\frac{\la}{3}\KK\Big( m(\p_m A)
+2\la A(\p_m A) \Big)\frac{1}{\Box}\big( (\p_n A^*)\Box A^*\big)\nn\\
&&+\frac{\la}{3}\KK\Big( m(\p_m A^*)
+ 2\la A^*(\p_m A^*) \Big)\frac{1}{\Box}\big( (\p_n A)\Box A\big)\nn\\
&&+\frac{\la}{12}\KKt\Big( mA^*+\la(A^*)^2 \Big)^3
+\frac{\la}{12}\KKtc\Big( mA+\la A^2\Big )^3\nn\\
&&-\frac{\la}{6}\KK\Big((mA+\la A^2)(\p_m A^*)(\p_n A^*)
+\big( mA^*+\la (A^*)^2\big)(\p_m A)(\p_n A)\Big)\nn\\
&& -\frac{\la}{3}\Big(\KK
\psi(\p_n\psi) - 2\KKi(\p_n\psi)\sigma^{ba}\psi\Big)(\p_m A)\nn\\
&& -\frac{\la}{3}\Big(\KK \bar\psi(\p_n\bar\psi) -
2\KKi\bar\psi\bar\sigma^{ab}(\p_n\bar\psi) \Big)(\p_mA^*) , \label{L2bezF}\\
{\cal L}_4 &=&\frac{\la^2}{24}\KK\KKt
\Big( mA^*+\la(A^*)^2 \Big) ^2(\p_m A^*)(\p_n A^*)\nn\\
&& + \frac{\la^2}{24}\KK\KKtc\Big( mA+\la A^2\Big)^2 (\p_m A)(\p_n A)\nn\\
&& - \frac{\la}{192}\KKt K^{*mn}K^*_{mn}(mA+\la A^2)(\Box A^*)^2\nn\\
&& -\frac{\la}{192}\KKt K^{*mn}K^*_{mn}(mA^*+\la (A^*)^2)(\Box A)^2\nn\\
&& -\frac{\la}{16}\KKt\KKtc \Big( mA+\la A^2\Big)^2\Big( mA^*+\la (A^*)^2\Big)^2\nn\\
&& -\frac{\la^2}{18}\KK K^{pb}K^{*q}_{\ \ b}\Big( (\p_m
A^*)(\p_n A^*)\Big) \frac{1}{\Box}\p_p\Big( (\p_q A)\Box A\Big)\nn\\
&&-\frac{\la^2}{18}\KK K^{pb}K^{*q}_{\ \ b}\Big( (\p_m
A)(\p_n A)\Big) \frac{1}{\Box}\p_p\Big( (\p_q A^*)\Box A^*\Big)\nn\\
&& - \frac{\la^2}{6}\KK\KKt(mA^*+\la (A^*)^2)\Big( m(\p_m A^*)
+ 2\la A^*(\p_m
A^*)\Big)\frac{1}{\Box}\Big( (\p_n A^*)\Box A^*\Big)\nn\\
&& - \frac{\la^2}{6}\KK\KKtc(mA+\la A^2)\Big( m(\p_m A)
+2\la A(\p_m A)\Big)\frac{1}{\Box}\Big( (\p_n A)\Box A\Big)\nn\\
&& -\frac{\la^2}{9}\KK K^p_{\ b}K^{*qb}\frac{1}{\Box}\p_m\Big((\p_n A)\Box
A\Big)\frac{1}{\Box}\p_p\Big( (\p_q A^*)\Box A^*\Big)\nn\\
&&-\frac{\la^2}{36}\KK K^p_{\ b}K^{*qb}(\p_m A)(\p_n A)(\p_p A^*)(\p_q A^*) . \label{L4bezF}
\end{eqnarray}

\section{Deformed Poincar\' e invariance}

Before commenting on the Lagrangian (\ref{LohneF}) we shall analyze the consequences
of the twist (\ref{twist}) on Poincar\' e symmetry. As in the case of the $\theta$-deformed space, the sub(Hopf)algebra of translations remains undeformed \cite{ponctwist}. Therefore we concentrate on the Lorentz transformations and first review some well known facts and formulas.

Under the infinitesimal Lorentz transformations the coordinates of the superspace transform as follows
\begin{eqnarray}
\delta_\omega x^m &=& \omega^m_{\ n}x^n, \label{deltax}\\
\delta_\omega \theta_\alpha &=& \omega^{mn}(\sigma_{mn})_\alpha^{\ \beta}\theta_\beta, \label{deltatheta}\\
\delta_\omega \bar{\theta}^{\dot{\alpha}} &=& \omega^{mn}
(\bar{\sigma}_{mn})^{\dot{\alpha}}_{\
\dot{\beta}}\bar{\theta}^{\dot{\beta}}, \label{deltabartheta}
\end{eqnarray}
where $\omega^{mn}= - \omega^{nm}$ are constant antisymmetric parameters.

The superfield $F$ (\ref{F}) is a scalar under the Lorentz transformations
\begin{equation}
F'(x', \theta', \bar{\theta}') = F(x, \theta, \bar{\theta}), \label{F'}
\end{equation}
or
\begin{eqnarray}
\delta_\omega F &=& F'(x, \theta, \bar{\theta}) - F(x, \theta, \bar{\theta})\nonumber\\
&=&\frac{1}{2}\omega^{mn}L_{mn}F(x, \theta, \bar{\theta})\nonumber\\
&=&\frac{1}{2}\omega^{mn}\Big( x_m\p_n - x_n\p_m
-(\sigma_{mn}\varepsilon)_{\alpha\beta}(\theta^\alpha\p^\beta + \theta^\beta\p^\alpha) \nonumber\\
&&-(\varepsilon\bar{\sigma}_{mn})_{\dot{\alpha}\dot{\beta}}(\bar{\theta}^{\dot{\alpha}}
\bar{\p}^{\dot{\beta}} + \bar{\theta}^{\dot{\beta}}\bar{\p}^{\dot{\alpha}}) \Big)
F(x, \theta, \bar{\theta}) . \label{Lmn}
\end{eqnarray}
To calculate the last line in (\ref{Lmn}) we used (\ref{deltax}),
(\ref{deltatheta}) and (\ref{deltabartheta}). Note that we use the same notation for transformations of coordinates and for variation of fields. The meaning should be clear from the context. Using the generators
$L_{mn}$ we can rewrite (\ref{deltax}), (\ref{deltatheta}) and
(\ref{deltabartheta}) in the following way
\begin{eqnarray}
\delta_\omega x^m &=& \omega^m_{\ \ n}x^n =
-\frac{1}{2}\omega^{rs}L_{rs}x^m, \label{deltax'}\\
\delta_\omega \theta_\alpha &=&
 \omega^{mn}(\sigma_{mn})_\alpha^{\ \beta}\theta_\beta =
-\frac{1}{2}\omega^{mn}L_{mn}\theta_\alpha, \label{deltatheta'}\\
\delta_\omega \bar{\theta}^{\dot{\alpha}} &=&
 \omega^{mn}(\bar{\sigma}_{mn})^{\dot{\alpha}}_{\ \dot{\beta}} \bar{\theta}^{\dot{\beta}}
= -\frac{1}{2}\omega^{mn}L_{mn} \bar{\theta}^{\dot{\alpha}} .\label{deltabartheta'}
\end{eqnarray}
Also,
\begin{equation}
\delta_\omega \theta^\alpha = -\omega^{mn}(\sigma_{mn})_\beta^{\ \alpha}\theta^\beta = -\frac{1}{2}\omega^{mn}L_{mn}\theta^\alpha.
 \label{deltathetauper}
\end{equation}

The Hopf algebra of the undeformed infinitesimal Lorentz transformations is given by
\begin{eqnarray}
\lb \delta_\omega, \delta_{\omega'} \rb &=& \delta_{\lb \omega, \omega' \rb}, \nonumber\\
\Delta (\delta_\omega) &=& \delta_\omega \otimes 1 + 1\otimes\delta_\omega , \nonumber\\
\varepsilon(\delta_\omega) &=& 0, \quad\quad S(\delta_\omega)=
-\delta_\omega .\label{undefHopfLor}
\end{eqnarray}
In terms of the generator $L_{mn}$ the coproduct reads
\begin{equation}
\Delta (L_{mn}) = L_{mn} \otimes 1 + 1\otimes L_{mn} . \label{undefcoprLmn}
\end{equation}

The twist $\cal{F}$ (\ref{twist}), when applied to
(\ref{undefHopfLor}), gives the Hopf algebra of the deformed Lorentz
transformations
\begin{eqnarray}
\lb \delta_\omega, \delta_{\omega'} \rb &=& \delta_{\lb \omega, \omega' \rb}, \nonumber\\
\Delta_{\cal{F}} (\delta_\omega) &=& {\cal F}
\Big( \delta_\omega \otimes 1 + 1\otimes\delta_\omega\Big){\cal F}^{-1}  \nonumber\\
&=& \delta_\omega \otimes 1 + 1\otimes\delta_\omega\nonumber\\
&& -\frac{1}{2}C^{\alpha\beta}\omega^{mn}\big(
\p_\alpha\otimes (\sigma_{mn}\varepsilon)_{\beta\gamma}\p^\gamma
+ (\sigma_{mn}\varepsilon)_{\alpha\gamma}\p^\gamma\otimes \p_\beta\big)\nonumber\\
&& - \frac{1}{2}\bar{C}_{\dot{\alpha}\dot{\beta}}\omega^{mn}\big(
\bar{\p}^{\dot{\alpha}}\otimes (\varepsilon\bar{\sigma}_{mn})_{\dot{\rho}\dot{\sigma}}
\varepsilon^{\dot{\sigma}\dot{\beta}}\bar{\p}^{\dot{\rho}}
+ (\varepsilon\bar{\sigma}_{mn})_{\dot{\rho}\dot{\sigma}}
\varepsilon^{\dot{\sigma}\dot{\alpha}}\bar{\p}^{\dot{\rho}} \otimes \bar{\p}^{\dot{\beta}}
,\nonumber\\
\varepsilon(\delta_\omega) &=& 0, \quad\quad S(\delta_\omega)=
-\delta_\omega .\label{defHopfLor}
\end{eqnarray}
The result for the deformed coproduct is the result
to all orders, as all higher order terms cancel since transformations
(\ref{Lmn}) are linear in coordinates. The algebra is
unchanged, but the comultiplication, leading to the deformed Leibniz rule,
changes. Form (\ref{defHopfLor}) one can see that the comultiplication for the deformed Lorentz transformations does not close in the algebra of Lorentz transformations, but in the bigger algebra with derivatives included. Therefore, we cannot speak about the deformed Lorentz symmetry but instead we have to work with the deformed Poincar\' e symmetry.

Now we give two examples for the application of the deformed Leibniz rule.

\begin{itemize}

\item
The $\star$-product of two Grassmanian coordinates should transform as in the undeformed case
\begin{eqnarray}
\delta_\omega (\theta^\alpha\star\theta^\beta) &=&
-\frac{1}{2}\omega^{mn}L_{mn} (\theta^\alpha\star\theta^\beta) \nonumber\\
&=&\frac{1}{2}\omega^{mn}
(\sigma_{mn}\varepsilon)_{\gamma\delta}(\theta^\gamma\p^\delta +
\theta^\delta\p^\gamma)
\big( \theta^\alpha\theta^\beta +\frac{1}{2}C^{\alpha\beta} \big) \nonumber\\
&=&-\omega^{mn}\Big( (\sigma_{mn})_\gamma^{\
\alpha}\theta^\gamma\theta_\beta + (\sigma_{mn})_\gamma^{\
\beta}\theta_\alpha\theta^\gamma \Big) .\label{pr1.1}
\end{eqnarray}
In the second line the $\star$-product is expanded and the definition of $L_{mn}$ given in (\ref{Lmn}) is used. Using the deformed coproduct on the other hand gives
\begin{eqnarray}
\delta_\omega (\theta^\alpha \star\theta^\beta) &=&
(\delta_\omega \theta^\alpha) \star \theta^\beta + \theta^\alpha \star (\delta_\omega \theta^\beta)\nonumber\\
&& -\frac{1}{2}C^{\rho\sigma}\omega^{mn}\Big(
(\p_\rho \theta^\alpha) \star (\sigma_{mn}\varepsilon)_{\sigma\gamma}(\p^\gamma \theta^\beta)
\nonumber\\
&& + (\sigma_{mn}\varepsilon)_{\rho\gamma}(\p^\gamma \theta^\alpha)
\star (\p_\sigma \theta^\beta) \Big)\nonumber\\
&=&-\omega^{mn}\Big( (\sigma_{mn})_\gamma^{\
\alpha}\theta^\gamma\theta_\beta + (\sigma_{mn})_\gamma^{\
\beta}\theta_\alpha\theta^\gamma \Big) .\label{pr1.2}
\end{eqnarray}
Comparing the results (\ref{pr1.1}) and (\ref{pr1.2}) we see that due to the deformed coproduct $\theta^\alpha \star\theta^\beta$
transforms as in the undeformed case. This type of calculation can also be done for $\star$-products of $\bar{\theta}$ coordinates with
the same conclusions.

\item When the $\star$-product of two chiral fields $\Phi_1$ and
$\Phi_2$ is expanded, the term
$C^{\alpha\beta}\psi_{1\alpha}\psi_{2\beta}$ appears. This term has to
transform as a scalar field under the deformed Poncar\' e
transformations, since it comes from $\Phi_1 \star\Phi_2$ which is a scalar field (using the deformed Leibniz rule of course).

Naively we have
\begin{eqnarray}
\delta_\omega(C^{\alpha\beta}\psi_{1\alpha}\psi_{2\beta})&=&
C^{\alpha\beta}\Big( (\delta_\omega \psi_{1\alpha})\psi_{2\beta}
+ \psi_{1\alpha}(\delta_\omega \psi_{2\beta}) \Big) \nonumber\\
&=& C^{\alpha\beta}\omega^{mn}\Big( (\sigma_{mn})_\alpha^{\
\gamma}\psi_{1\gamma}\psi_{2\beta}
+ (\sigma_{mn})_\beta^{\ \gamma}\psi_{1\alpha}\psi_{2\gamma}
+\frac{1}{2} (x_m\p_n - x_n\p_m)(\psi_{1\alpha}\psi_{2\beta}) \Big) \nonumber\\
&\neq& \frac{1}{2}\omega^{mn}L_{mn} \big(
C^{\alpha\beta}\psi_{1\alpha}\psi_{2\beta}\big) ,\label{pr2.1}
\end{eqnarray}
with $L_{mn}$ defined in (\ref{Lmn}). The equality sign in the last line can be achieved by transforming the fields $\psi_{1\alpha}$ and $\psi_{2\beta}$ not as spinor fields (as it was done in (\ref{pr2.1})) but as scalar fields. The reason for this is that indices $\alpha$ and $\beta$ are contracted with indices on $C^{\alpha\beta}$. Namely, the twist $\cal{F}$ (\ref{twist}) is a globally defined object \cite{paoloproc}. Therefore, under the transformations (\ref{deltatheta}) and (\ref{deltabartheta}) the derivatives $\p$ and $\bar{\p}$ appearing in $\cal{F}$ transform in the following way
\begin{equation}
\delta_\omega \p_\alpha = \delta_\omega \bar{\p}_{\dot{\alpha}} = 0 .\label{pr2.5}
\end{equation}
Also, $C^{\alpha\beta}$ and $\bar{C}^{\dot{\alpha}\dot{\beta}}$ (being complex constants) do not transform. Therefore, all indices contracted with $C^{\alpha\beta}$ and $\bar{C}^{\dot{\alpha}\dot{\beta}}$ should be understood as scalar (non-transforming) indices.

To convince ourselves that this is the right way of thinking let us rewrite $C^{\alpha\beta}\psi_{1\alpha}\psi_{2\beta}$ by using the $\star$-product and then use the deformed Leibniz rule to transform it
\begin{eqnarray}
C^{\alpha\beta}\psi_{1\alpha}\psi_{2\beta}&=&
-2\theta^\alpha\psi_{1\alpha}\star \theta^\beta\psi_{2\beta}
- \theta\theta\psi_1^\alpha\psi_{2\alpha} \nonumber\\
&=& -2\theta^\alpha\psi_{1\alpha}\star \theta^\beta\psi_{2\beta}
-(\theta^\alpha \star \theta_\alpha) \psi_1^\beta\psi_{2\beta} \nonumber\\
\delta_\omega(C^{\alpha\beta}\psi_{1\alpha}\psi_{2\beta})&=&
-2\delta_\omega\big( \theta^\alpha\psi_{1\alpha}\star
\theta^\beta\psi_{2\beta}\big) - \delta_\omega \big(
(\theta^\alpha \star \theta_\alpha )\psi_1^\beta\psi_{2\beta} \big).\label{pr2.2}
\end{eqnarray}
Note that $\psi_1^\beta \star \psi_{2\beta} = \psi_1^\beta \psi_{2\beta}$. Also note that $\delta_\omega$ in this example is the variation of a field as in (\ref{Lmn}). Therefore
\begin{eqnarray}
\delta_\omega(\theta^\alpha\psi_{1\alpha}) &=& \theta^\alpha\delta_\omega(\psi_{1\alpha}) \nonumber\\
&=& \frac{1}{2}\omega^{mn}L_{mn}\big( \theta^\alpha\psi_{1\alpha} \big). \nonumber
\end{eqnarray}

Let us calculate the transformation of the first term in (\ref{pr2.2})
\begin{eqnarray}
\delta_\omega\big( \theta^\alpha\psi_{1\alpha}\star
\theta^\beta\psi_{2\beta}\big) &=& \big( \delta_\omega
(\theta^\alpha\psi_{1\alpha})\big) \star
(\theta^\beta\psi_{2\beta}) +
(\theta^\alpha\psi_{1\alpha}) \star \big( \delta_\omega (\theta^\beta\psi_{2\beta}) \big) \nonumber\\
&& -\frac{1}{2}C^{\rho\sigma}\omega^{mn}\Big(
\big( \p_\rho (\theta^\alpha\psi_{1\alpha}) \big) \star (\sigma_{mn}\varepsilon)_{\sigma\gamma}
\big( \p^\gamma (\theta^\beta \psi_{2\beta})\big) \nonumber\\
&&+ (\sigma_{mn}\varepsilon)_{\rho\gamma}\big( \p^\gamma (\theta^\alpha\psi_{1\alpha}) \big)
\star \big(\p_\sigma (\theta^\beta \psi_{2\beta})\big) \Big)\nonumber\\
&=&\frac{1}{2}\omega^{mn}L_{mn} \Big(
\theta^\alpha\psi_{1\alpha}\star \theta^\beta\psi_{2\beta} \Big)
.\label{pr2.3}
\end{eqnarray}
We conclude that $\theta^\alpha\psi_{1\alpha}\star
\theta^\beta\psi_{2\beta}$ is a scalar field. Calculation  similar to this shows that $(\theta^\alpha \star \theta_\alpha )\psi_1^\beta\psi_{2\beta}$ is also a scalar field. Thus, we have demonstrated that $C^{\alpha\beta}\psi_{1\alpha}\psi_{2\beta}$ really
transforms as a scalar field.

\end{itemize}

\section{Conclusions and outlook}

The Lagrangian (\ref{LohneF}) is the final result of this paper.
By construction this Lagrangian is covariant under the deformed SUSY
transformations (\ref{defsusytr}) and leads to a deformed SUSY
invariant action. Note that it is the deformed Leibniz rule which enables this construction. No new fields appear in the action, the deformation is present
only trough some new interaction terms. The deformation parameter plays
the role of a coupling constant and in the limit $C\to 0$ the
undeformed theory is obtained. If this leads to some new physics remains to be understood by further analysis of our model.

At the moment we are interested in the renormalization
properties of (\ref{LohneF}), first of all in the cancellation of the quadratic
divergences. Let us comment that it is possible to choose a
specific type of deformation, such that it leads to
$K^{ab}K_{ab}=K^{*ab}K^*_{ab}=0$. This choice takes the
$H^3$ term in (\ref{Lincompfields}) to zero and
simplifies calculations drastically. More important,
renormalization properties of our model might turn out to be
better with this choice.

One should analyze microcausality of our theory since a
non-local interaction term appears in the action. Also, the
construction of gauge theories on this deformed superspace is
planed for future research.

Concerning different types of deformation, we also analyzed a
model with ${\cal F} = e^{\frac{1}{2}C^{\alpha\beta}D_\alpha
\otimes D_\beta}$ which leads to the deformation discussed in
\cite{Ferrara}. Comments on this work are planed for the next
publication.

\vspace*{0.5cm}
\begin{flushleft}
{\Large {\bf Acknowledgments}}
\end{flushleft}

Authors would like to thank Paolo Aschieri for the interesting discussions and the many insights. Also, M.D. and V.R. would like to thank II Institute for
Theoretical Physics and DESY, Hamburg for their hospitality during
their stay in December 2006 and July 2007. M.D. also thanks
Quantum Geometry and Quantum Gravity Research Networking Programme
of the European Science Foundation for their financial support
during one short stay in Alessandria, Italy where part of this
work was completed. The work of M.D. and V.R. is supported by the
project $141036$ of the Serbian Ministry of Science.

\vspace*{0.5cm}
\begin{flushleft}
{\Large {\bf Final remark}}
\end{flushleft}

This work was completed in July (all except the section 7, for which only the idea was given at that time) when all the authors enjoyed the
hospitality of II Institute for Theoretical Physics and DESY,
Hamburg. Sadly and unexpectedly Julius Wess passed away in
August. We were left with piles of handwritten calculations
and no paper written. Much more important, we stayed without
Julius's enormous knowledge and experience, his ideas,
encouragement, support and his friendship. He enjoyed this work
very much and we only hope that he would not object the way we
wrote it up too much.

\end{document}